# Wavelet Denoised-ResNet CNN and LightGBM Method to Predict Forex Rate of Change

Yiqi Zhao
School of Computer Science, The University of Sydney, NSW Australia
yzha9512@uni.sydney.edu.au

Matloob Khushi
School of Computer Science, The University of Sydney, NSW Australia
matloob.khushi@sydney.edu.au

*Abstract*— Foreign Exchange (Forex) is the largest financial market in the world. The daily trading volume of the Forex market is much higher than that of stock and futures markets. Therefore, it is of great significance for investors to establish a foreign exchange forecast model. In this paper, we propose a Wavelet Denoised-ResNet with LightGBM model to predict the rate of change of Forex price after five time intervals to allow enough time to execute trades. All the prices are denoised by wavelet transform, and a matrix of 30 time intervals is formed by calculating technical indicators. Image features are obtained by feeding the maxtrix into a ResNet. Finally, the technical indicators and image features are fed to LightGBM. Our experiments on 5-minutes USDJPY demonstrate that the model outperforms baseline modles with MAE: $0.240977\times10^{-3}$ MSE: $0.156\times10^{-6}$ and RMSE: $0.395185\times10^{-3}$. An accurate price prediction after 25 minutes in future provides a window of opportunity for hedge funds algorithm trading. The code is available from https://mkhushi.github.io/

*Keywords— Forex, wavelet, ResNet, LightGBM*

.INTRODUCTION

Foreign Exchange (Forex) is the largest financial market in the world. The daily trading volume of the Forex market is more than 3 trillion US dollars, far higher than that of stock and futures markets [1]. The trend of Forex price has a direct impact on the economic interest of investors, and also reflects the national macroeconomic policy, so it has been widely concerned in society. Due to the volatility of Forex prices, investors will inevitably encounter investment risks[2]. Modeling the stock and Forex market and forecasting the price and trend is not only of attractive application value but also of great significance to economic development and financial construction.

Forex is 24 hours market, hence the models could be continuously updated and iterated, and a continuous prediction of Forex price trends can be realized. Identifying timely trend is important to guide investors and hedge funds to accurately grasp the trading opportunity and reduce investment risk [3]. The higher the accuracy of forecasting the future Forex price trend, the higher the probability of investors' profit.

Wavelet analysis, came from Fourier analysis, is a signal processing tool developed in recent years. Wavelet only has a non-zero value in a very limited range, rather than being endless like sine and cosine waves [4]. Wavelets can be shifted back and forth along the time axis, and can also be extended and compressed in proportion to obtain low-frequency and high-frequency wavelets. The constructed wavelet function can be used to filter or compress signals, so as to extract useful signals from noisy signals. It can be used in many fields, such as astronomy, medical image, and computer vision [5]. In this paper, this method is used for processing of open, high, low, and close prices (OHLC) of a time interval.

Deep learning has proved itself a breakthrough technological in many fields such as image, voice, natural speech processing, reaching the level of practical application, and also began to be applied to the financial field [6]. To capture the price trend we treated the Forex precises as an image matrix, extracted image features using convolution neural network (CNN) following by further processing by Light gradient boosting machine (LightGBM).

LightGBM is a boosting framework in ensemble learning, which uses a decision tree as the base classifier of learning algorithms [7]. The boosting method uses the serial training function to train the base classifiers, and each classifier depends on each other. Its basic idea is to stack each base classifier layer by layer and upgrade the weak classifier to the strong one. In each training iteration, the higher weights are given to the previous training error samples, and gradually stack the classifier to a complex and powerful integrated classifier. During the prediction, the final prediction results are obtained by weighting the classifier results of each layer [8]. In this paper, LightGBM will be used to replace the softmax layer of the ResNet CNN and the technical indicators will also be added as the input to get the final result.

In summary, the proposed model consists of three core methods. The steps are as following and the structure is shown in Figure 1.

1. All the open, high, low, close (OHLC) price are denoised through the wavelet denoising, then the technical indicators using the denoised value are calculated.

2. Matrix of of 30 time intervals fed into ResNet CNN for training and feature extraction

3. The extracted matrix (image) features fed into LightGBM together with technical indicators for training to predict the rate of change after fifth time interval.

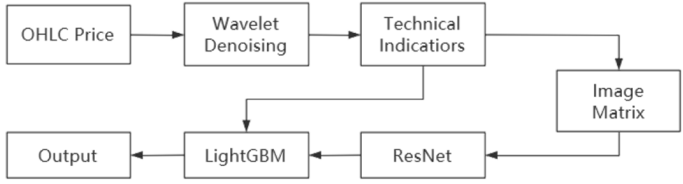

Fig. 1. The overall structure of the proposed model

.RELATED WORK

*2.1 The application of Wavelet Denoising*

Wavelet denoising is an effective method to separate signals from noise in many practical applications and it can be applied to many fields. In the field of image processing, the denoising method based on wavelet will introduce less smoothness and retain the sharpness of the image, and the original shape of the active region [9]. As for the infrared spectra, when the wavelet denoising method is applied to pure infrared spectra with various added levels of homo- and heteroscedastic noise, it will have a better root mean square errors and visual quality compared with pure spectra [10].

In the financial field, such as stock and Forex, wavelet denoising also plays an important role in the construction of the prediction model. When using this method to denoise the financial data, and then put it into the LSTM model to predict the data, it will often get better results [11]. When using the Saudi stock index (SSI) time series as the carrier for prediction and analysis, the application of wavelet denoising in its time series plays an important role. The results show that the information of SSI can be fully utilized by using wavelet transform to decompose SSI into multiple sequences with different resolutions [12].

*2.2 Deep Learning in Stock/Forex Market Prediction*

The method of deep learning is one of the most advanced and promising methods to predict Forex or stock and other financial data[13]. The feature extraction of financial data is an important problem in the field of market prediction, and CNN is one of the best tools for image feature selection and price trend prediction. A CNN based framework performed better than the existing baseline algorithm by obtaining features from the collected data set (S&P 500, NASDAQ, DJI, NYSE, and RUSSELL) to predict the future price trend [14].

LSTM has a good performance in dealing with the time series model, so it is also a hot topic to apply LSTM to the prediction of Forex or stock price. LSTM can support any size of the time step, and there is no vanishing gradient problem. When a bidirectional or stacked LSTM prediction model is used, it performs better than the baseline shallow LSTM model [15].

A C-RNN forecasting method for Forex time series data based on deep-Recurrent Neural Network (RNN) and deep-Convolutional Neural Network (CNN) can further improve the prediction accuracy of a deep learning algorithm for the time series data of exchange rate. Meanwhile, it can fully exploit the Spatio-temporal characteristics of forex time series data based on the data-driven method [16].

*2.3 The application of LightGBM*

LightGBM is one of the latest decision model algorithm proposed by researchers at Microsoft in 2017. It has not been widely used in the financial field. In this paper, LightGBM is innovatively applied to the field of foreign exchange forecasts.

LightGBM can be used as a powerful tool for the recognition and classification of breast cancer miRNA targets. Compared with the performance of random forest (RF) and extreme gradient boosting (XGBoost), it has better performance in terms of precision and logic loss [17]. For predicting cryptocurrency price trend LightGBM has also shown to perform well [8].

.METHODOLOGIES

*3.1 Wavelet Domain Denoising and normalization*

The idea of wavelet threshold denoising is that after the signal is transformed by wavelet, the wavelet coefficients generated by the signal contain the important information of the signal[18]. After the signal is decomposed by wavelet, the wavelet coefficients of the signal are larger, the wavelet coefficients of the noise are smaller, and the wavelet coefficients of the noise are smaller than the wavelet coefficients of the signal. By selecting a suitable threshold, the wavelet coefficients larger than the threshold are considered to have signals. If it is less than the threshold value, it is considered that it is generated by noise and set to zero to achieve the purpose of noise reduction [19].

From the point of view of signal science, wavelet denoising is a problem of signal filtering. Although wavelet denoising can be regarded as a low-pass filter to a large extent, it is superior to traditional low-pass filter in this respect because it can retain signal characteristics successfully after denoising [20]. It can be seen that wavelet denoising is the synthesis of feature extraction and low-pass filtering, and its flow chart is as shown in figure2.

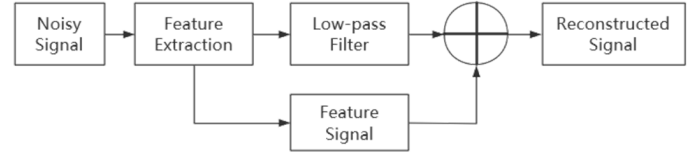

Fig. 2. The flow chart of wavelet denoising method

A noisy model can be represented as follows:

$$S(k) = f(k) + \varepsilon * e(k) \quad k = 0.1 \ldots \ldots n\text{-}1 \quad (1)$$

Among them, $f(k)$ is a useful signal, $s(k)$ is a noisy signal, $e(k)$ is noise, and ε is the standard deviation of the noise coefficient.

In this paper, all the OHLC prices are denoised through this method. After the wavelet threshold denoising, the technical indicators are calculated through the denoised OHLC. At this time, the data need to be normalized through the following formula to avoid the scaling effect.

$$x_{\text{norm}} = \frac{x_t - x_{min}}{x_{max} - x_{min}} \quad (2)$$

The Fig. 3. shows the original technical indicators image, the Fig. 4. shows the denoised technical indicators image and the Fig 5. shows the denoised and normalized technical indicators image.

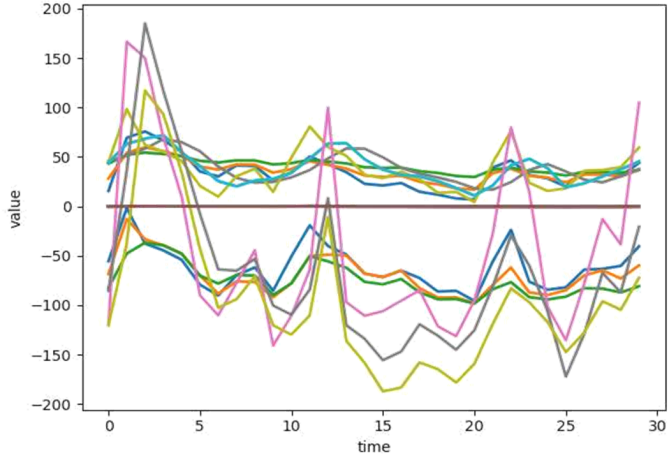

Fig. 3. The original technical indicators image

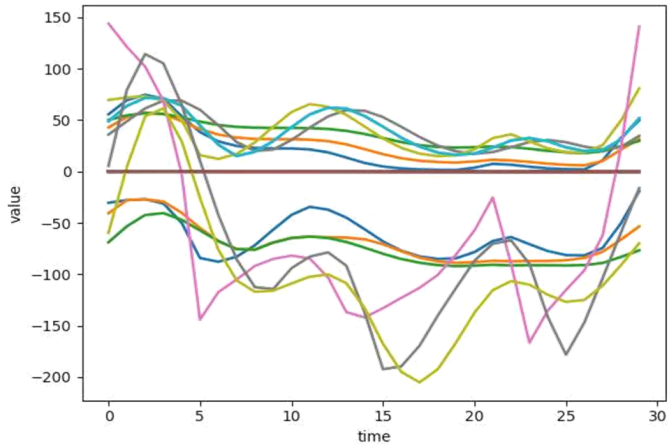

Fig. 4. The denoised technical indicators image

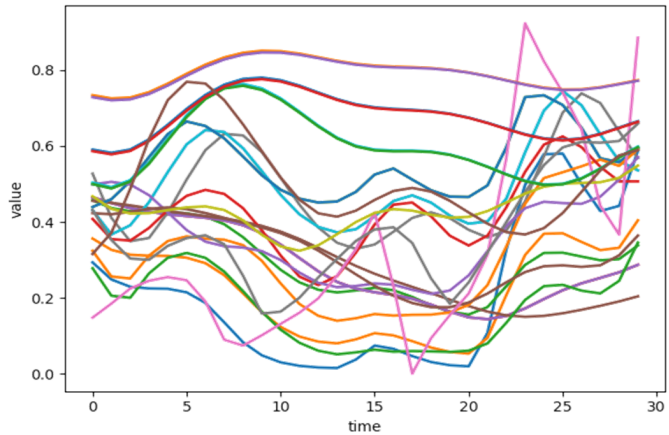

Fig. 5. The denoised and normalized technical indicators image

*3.2 ResNet CNN*

The traditional convolution network or the all connected network will have some problems, such as information loss when the features are transmitted. At the same time, it will cause the gradient to disappear or the gradient to explore, which makes the deep network unable to train. ResNet solves this problem to a certain extent by bypassing the input information directly to the output to protect the integrity of the information. The whole network only needs to learn the difference between the input and output to simplify the learning objectives and difficulties.

In this paper, the residual network is composed of a series of residual blocks. The residual block is divided into two parts: the direct mapping part and the residual part. $(x_l)$ is a direct mapping and $F(x_l, W_l)$ is the residual part. The transmission of the whole network information should be in a smooth channel, which requires that both and $f$ must be identity mapping [21].

$$y_l = (x_l) + F(x_l, W_l) \qquad (3)$$

$$x_{l+1} = f(y_l) \qquad (4)$$

Therefore, the formula can be combined to get:

$$x_{l+1} = x_l + F(x_l, W_l) \qquad (5)$$

The xl can be divided into the sum of the output of the previous module and the l layer residual module, so the loop recursion results in the following formula:

$$x_{l+2} = x_{l+1} + F(x_{l+1}, W_{l+1})$$
$$= x_l + F(x_l, W_l) + F(x_{l+1}, W_{l+1}) \qquad (6)$$

$$x_L = x_l + \sum_{i=l}^{L-1} F(x_i, W_i) \qquad (7)$$

For the output of L layer, it can be regarded as the superposition of the output $x_l$ of $i$ layer which is the layer before any L layer and the output of the intermediate residual block. The input of the intermediate residual block also changes with $i$, so each residual block also will be affected by $x_l$ the output of the L layer. Therefore, the whole network will be residual fashion. Any layer and any layer before the layer can be regarded as a residual module, which can ensure the smooth forward propagation of the whole network [22].

The backpropagation formula of ResNet is as follows:

$$\frac{\partial_l}{\partial_{xl}} = \frac{\partial_l}{\partial_{xL}} \frac{\partial_{xL}}{\partial_{xl}} = \frac{\partial_l}{\partial_{xL}} \left(1 + \frac{\partial}{\partial_{xl}} \sum_{i=l}^{L-1} F(x_i, W_i)\right) \qquad (8)$$

The gradient of X for any layer is composed of two parts, one part is directly conducted by L layer without any attenuation and change, which ensures the effectiveness of gradient propagation, the other part is changed from cumulative multiplication of chain rule to accumulation, so it has better stability [23].

In this paper using a moving window for every 30 time intervals all the technical indicators that have been wavelet denoised and normalized forming a 30 x 30 matrix. The matrix processed using ResNet for training. In the proposed model, the setting and structure of the residual block are as shown in the Fig. 6 where c is the channel number of the output, k is kernel size, p is padding size, s is strides. The overall ResNet has 50 layers which includes 16 residual blocks in total. The dimension of the input image is the 1×30×30 and the output is 1024 features extracted from the network. The setting and structure of the whole network are as shown in the Figure 7 where c is the channel number of the output, k is kernel size, p is padding size, s is strides,

in_planes, and planes are described in the structure of the residual block.

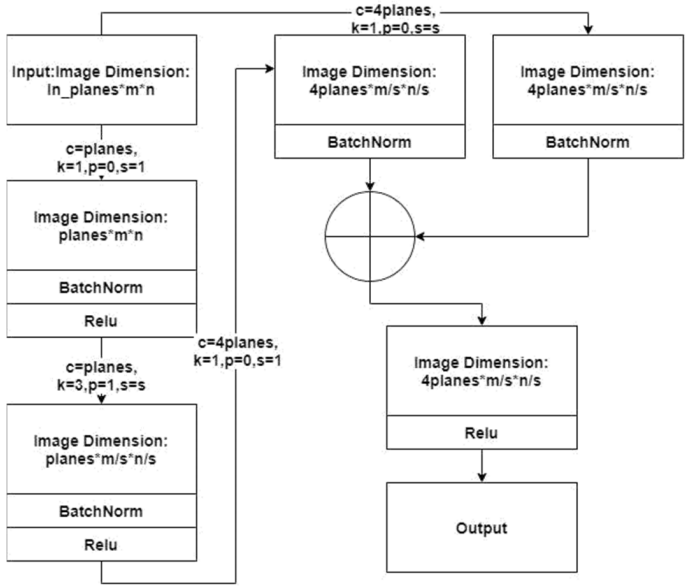

Fig. 6. The structure of the residual block

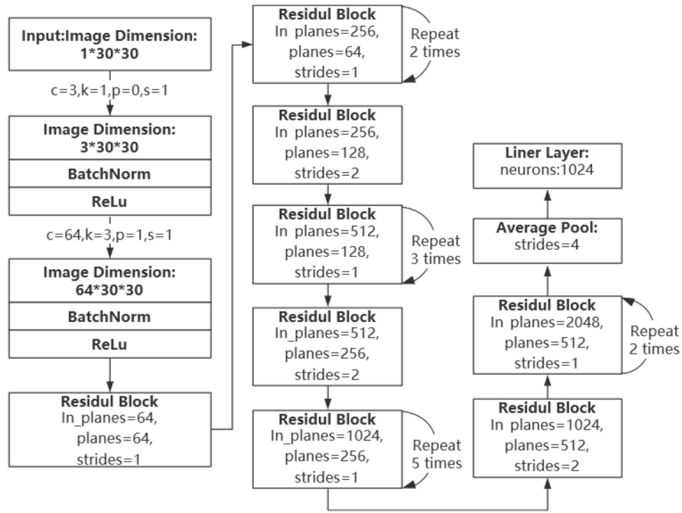

Fig. 7. The structure of the whole ResNet

*3.3 LightGBM*

The feature extracted from the ResNet and technical indicators were processed by LightGBM. In this step, the final prediction results were obtained by the direct data features and image features, which made the whole model very robust.

In the implementation of LightGBM, not only data sampling but also feature sampling is carried out, which further improved the training speed of the model [24]. The feature sampling is different from the general feature sampling, which is to bind mutually exclusive features together to reduce the feature dimension [7]. The main idea is that the high dimension data is usually sparse (such as one-hot coding), which makes it possible to design a nearly lossless method to reduce the number of effective features. In particular, many features are mutually exclusive in sparse feature space (for example, non-zero values rarely occur at the same time). This makes it is possible to safely bind mutually exclusive features together to form a feature, thus reducing the feature dimension. And histograms based methods are used to bind mutually exclusive features together [25, 26].

The flow of the application of LightGBM algorithm is as follows:

1. Algorithm: Light GBM regression prediction
2. Input: high-level image feature data extracted by ResNet and the technical indicators
3. Output: The rate of change prediction for the 5$^{th}$ time interval.
4. Initialize m decision trees; the weight of training samples is 1 / m
5. Training weak classifier $f(x)$
6. According to the training error, the weight of the current weak classifier $f(x)$ is determined
7. When the maximum number of iterations is not reached, return to step 5 to continue training. When the maximum number of iterations is reached, enter step 8;
8. The final classifier is obtained as the formula :

$$f_m(x) = \partial_0 f_0(x) + \partial_1 f_1(x) + \partial_2 f_2(x) + \ldots + \partial_i f_i(x) + \cdots + \partial_m f_m(x) \quad (9)$$

9. The combined base classifier is a strong classifier to predict the result.

Among them, *m* is the number of algorithm iterations, *i* is the $i^{th}$ generation of iterations; where, $0 \leq i \leq m$ and it also represents $i^{th}$ trained classifiers of the whole iterations.

. EXPERIMENT SETUP

*4.1 Data Description*

We collected 5-minutes data of USDJPY currency pair which is one a major Forex currency pairs from 2019-01-01 to 2020-06-10 consisting 107,236 data rows in total. We used 80% of initial data, and the remaining 20% is used for testing.

As explained earliar the OHLC prices are denoised by wavelet, and the technical indicators are calculated from the denoised prices. The technical indicators used in this paper are mainly divided into two types, as follows:

1. Momentum Indicators: RSI5, RSI10, RSI20, MACD, MACDhist, MACDsignal, Slowk, Slowd, Fastk, Fastd, WR5, WR10, WR20, ROC5, ROC10, ROC20, CCI5, CCI10, CCI20

2. Volume Indicators: ATR5, ATR10, ATR20 , NATR5, NATR10, NATR20, TRANGE

As for the label of the whole proposed model, it is as shown below where $y_i$ is the close price (before wavelet denoising) of the current bar, $y_{i+5}$ is the close price (before wavelet denoising) of 5$^{th}$ bar, $label_i$ is the rate of change calculated as below:

$$label_i = (y_{i+5} - y_i)/yi \quad (10)$$

## 4.2 Evaluation metrics

Three evaluation metrics are used in this paper, they are MAE, RMSE, MSE respectively.

MAE: Mean Absolute Error, it is the average of absolute errors. It can better reflect the actual situation of the predicted value error.

$$MAE(X, ) = \frac{1}{m}\sum_{i=1}^{m} | (x_i) - y_i | \quad (11)$$

Where $(x_i)$ is the prediction value at the time $i$, $y_i$ is the and ground truth value at time t and m is the total number of the train or test data.

RMSE: Root Mean Square Error, it is a measure of the deviation between the observed value and the real value and it is often used as a standard for the prediction results of machine learning models

$$RMSE(X, ) = \sqrt{\frac{1}{m}\sum_{i=1}^{m} ( (x_i) - y_i )^2} \quad (12)$$

Where $(x_i)$ is the prediction value at the time i, $y_i$ is the and ground truth value at time t and m is the total number of the train or test data.

MSE: Mean Square Error, it is the square of the difference between the real value and the predicted value, and then the sum is averaged. It is easy to find derivative by square form, so it is often used as a loss function of linear regression.

$$MSE = \frac{1}{m}\sum_{i=1}^{m}(y_i - (x_i))^2 \quad (13)$$

Where $(x_i)$ is the prediction value at the time i, $y_i$ is the and ground truth value at time t and m is the total number of the train or test data.

## 4.3 Parameter Settings

In this section, we describe the parameters of the wavelet denoising method, ResNet and LightGBM.

Wavelet Denoising: The sym15 wavelet function is selected and the high-frequency coefficients of two levels are set to zero.

ResNet: All the hyper-parameters of the ResNet are shown in the Table I.

TABLE I. THE HYPER-PARAMETERS OF RESNET

| Parameter Name | Parameter Value |
| --- | --- |
| Number of ResNet layers | 50 |
| Number of residual block layers | 3 |
| Dimension of input | 1×30×30 |
| Dimension of output | 1024 |
| Activation Function | Relu |
| Batch Size | 128 |
| Learning Rate | 0.001 |
| Epochs | 100 |

LightGBM: All the hyper-parameters of the LightGBM are shown in the Table II.

TABLE II. THE HYPER-PARAMETERS OF LIGHTGBM

| Parameter name | Parameter value |
| --- | --- |
| Num_leaves | 38 |
| Min_data_in_leaf | 50 |
| Objective | Regression |
| Max_depth | -1 |
| Learning Rate | 0.02 |
| Min_sum_hessian_in_leaf | 6 |
| Boosting | GBDT |
| Feature_fraction | 0.9 |
| Bagging_freq | 1 |
| Bagging_fraction | 0.7 |
| Bagging_seed | 11 |
| Lambda_l1 | 0.1 |
| Verbosity | -1 |
| Nthread | 4 |
| Random_state | 2019 |

## RESULTS AND DISCUSSION

During the experiments stage, in order to verify the effectiveness of wavelet denoising, ResNet, and LightGBM in predicting the rate of change of Forex price, we performed several experimental models, which are respectively:

1. Traditional CNN (TC) model: In this model, the input image of Forex technical indicators do not undergo wavelet denoising, and it is processed by the traditional structure of CNN for training, and the final training results is obtained without LightGBM.

2. Denoised Traditional CNN (DTC) model: In this model, the input image of Forex technical indicators undergoes wavelet denoising, and is processed by traditional structure of CNN for training, and the final training results obtained without LightGBM.

3. Traditional CNN LightGBM (TCL) model: In this model, the input image of Forex technical indicators do not undergo wavelet denoising, and it is processed by traditional structure of CNN for training, and the final training results is obtained with LightGBM.

4. Denoised Traditional CNN LightGBM (DTCL) model: In this model, the input image of Forex technical indicators undergoes wavelet denoising, and it is processed by the traditional structure of CNN for training, and the final training results are obtained with LightGBM.

5. ResNet (RN) model: In this model, the input image of Forex technical indicators does not undergo wavelet denoising, and it is processed by the ResNet for training, and the final training results are obtained without LightGBM.

6. Denoised ResNet (DRN) model: In this model, the input image of Forex technical indicators undergoes wavelet denoising, and it is processed by the ResNet for training, and the final training results were obtained without LightGBM.

7. ResNet LightGBM (RNL) model: In this model, the input image of Forex technical indicators does not undergo wavelet denoising, and it is processed by the ResNet for training, and the final training results is obtained with LightGBM.

8. Denoised ResNet LightGBM (DRNL) model: It is the proposed model. In this model, the input image of Forex technical indicators undergoes wavelet denoising, and processed by the ResNet for training, and the final training results were obtained by LightGBM.

Table III shows the test results of evaluation metrics of the above models.

TABLE III. THE COMPARISON OF THE EXPERIMENTAL RESULTS

| MODEL | MAE(X$10^{-3}$) | MSE(X$10^{-6}$) | RMSE(X$10^{-3}$) |
|---|---|---|---|
| TC | 0.585036 | 0.680 | 0.824666 |
| DTC | 0.400671 | 0.468 | 0.683787 |
| TCL | 0.361741 | 0.382 | 0.617769 |
| DTCL | 0.28278 | 0.215 | 0.463204 |
| RN | 0.394484 | 0.508 | 0.713024 |
| DRN | 0.372901 | 0.424 | 0.650970 |
| RNL | 0.362895 | 0.365 | 0.603879 |
| **DRNL*** | **0.240977** | **0.156** | **0.395185** |

According to the results shown above in the Table III, it is identified when using the traditional CNN architecture, DTC performs best with the lowest MAE, MSE, and RMSE. Compared with the TC model, the DTC and TCL model both have an improvement which means the wavelet-denoising method and the LightGBM method can help the CNN model to improve the performance when predicting the Forex rate of change. DTCL model is the second best has a significant improvement compared with other models. It shows that the wavelet-denoising and the LightGBM method improves the performance of a CNN model.

Similarly, when using the ResNet architecture, DRNL performs best with the lowest MAE, MSE, and RMSE. Compared with the other RN models, the DRN and TRN model both show improvements which means the wavelet-denoising and the LightGBM method can help the ResNet model to improve the performance when predicting the forex change rate. Meanwhile, the DRNL model has a significant improvement compared with other models, it confirms that the wavelet-denoising method and the LightGBM method improve the performance of the ResNet model.

From these two perspectives, they proved the feasibility of wavelet-denoising and LightGBM, and the ability to improve the performance of any CNN model. Our propsed DRNL model is the best among the eight models for all three evaluation metrics MAE, MSE or RMSE. It will show that the ResNet model can outperform the traditional CNN model when predicting the Forex rate of change of the price.

The Figure 8 shows the comparison of the real price (ground truth) and predicted value on a typical trading day of 24 hours on 22 April 2020. The predicted and real (ground trouth) prices pretty much overlay each other. Figure 9 shows a zoomed comparison of the real and predicted value for 8 data points (35 minutes). As this could be seen that predicted and real prices are very close to each other making the model really useful for practical trading.

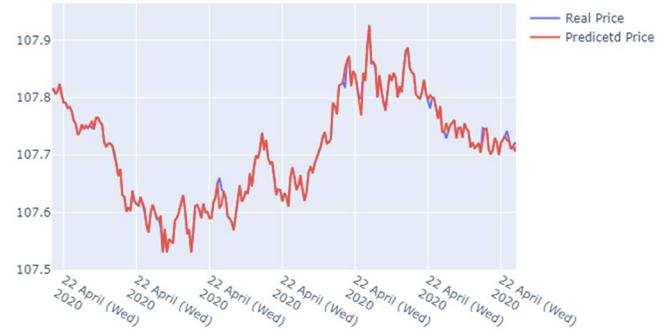

Fig. 8. Comparison of real (ground truth) and predicted prices on typical trading day for 24 hours.

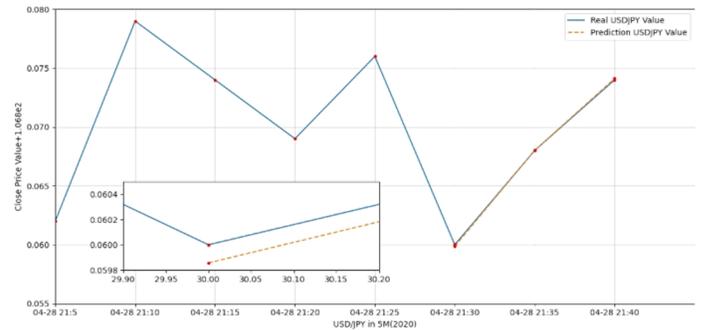

Fig. 9. Comparison in a precise dimension

. CONCLUSION

In this paper, a wavelet denoised ResNet-LightGBM model is proposed to predict the rate of change of price after fifth time interval. This means for 5 minutes data the model can precisely predict the price in future after 25 minutes. It has been shown that the ResNet has a better prediction ability than the traditional CNN network, by the ensambling of wavelet-denoising and LightGBM improved model performance significantly. Being able to predit price 5 data points ahead is very practical useful for investors and traders can used in hedge funds trading strategy.

Although this paper has achieved satisfactory results and has excellent practical significance, there are still some limitations, as shown below:

As a limitation and future direction we have only performed experiments on 5-minutes data, in future, we aim test the model on lower and higher time frames such as 15 minutes, 30 minutes, or 1 hour data.

REFERENCE

1. Nassirtoussi, A.K., T.Y. Wah, and D.N.C. Ling, A novel FOREX prediction methodology based on fundamental data.

1. African Journal of Business Management, 2013. 5(20): p. 8322-8330.
2. Z. Zhang and M. Khushi, "GA-MSSR: Genetic Algorithm Maximizing Sharpe and Sterling Ratio Method for RoboTrading," in 2020 International Joint Conference on Neural Networks (IJCNN), 19-24 July 2020
3. Vyklyuk, Y., D. Vukovic, and A. Jovanovic, Forex prediction with neural network: usd/eur currency pair. Актуальні проблеми економіки, 2013(10): p. 261-273.
4. Jaiswal, A., J. Upadhyay, and A. Somkuwar, Image denoising and quality measurements by using filtering and wavelet based techniques. AEU-International Journal of Electronics and Communications, 2014. 68(8): p. 699-705.
5. Srivastava, M., C.L. Anderson, and J.H. Freed, A new wavelet denoising method for selecting decomposition levels and noise thresholds. IEEE access, 2016. 4: p. 3862-3877.
6. Chong, E., C. Han, and F.C. Park, Deep learning networks for stock market analysis and prediction: Methodology, data representations, and case studies. Expert Systems with Applications, 2017. 83: p. 187-205.
7. Ke, G., et al. Lightgbm: A highly efficient gradient boosting decision tree. in Advances in neural information processing systems. 2017.
8. Sun, X., M. Liu, and Z. Sima, A novel cryptocurrency price trend forecasting model based on LightGBM. Finance Research Letters, 2020. 32: p. 101084.
9. Wink, A.M. and J.B. Roerdink, Denoising functional MR images: a comparison of wavelet denoising and Gaussian smoothing. IEEE transactions on medical imaging, 2004. 23(3): p. 374-387.
10. Alsberg, B.K., et al., Wavelet denoising of infrared spectra. Analyst, 1997. 122(7): p. 645-652.
11. Li, Z. and V. Tam. Combining the real-time wavelet denoising and long-short-term-memory neural network for predicting stock indexes. in 2017 IEEE Symposium Series on Computational Intelligence (SSCI). 2017. IEEE.
12. Alrumaih, R.M. and M.A. Al-Fawzan, Time Series Forecasting Using Wavelet Denoising an Application to Saudi Stock Index. Journal of King Saud University-Engineering Sciences, 2002. 14(2): p. 221-233.
13. T. L. Meng and M. Khushi, "Reinforcement Learning in Financial Markets," Data, vol. 4, no. 3, 2019
14. Hoseinzade, E. and S. Haratizadeh, CNNpred: CNN-based stock market prediction using a diverse set of variables. Expert Systems with Applications, 2019. 129: p. 273-285.
15. Althelaya, K.A., E.-S.M. El-Alfy, and S. Mohammed. Evaluation of bidirectional lstm for short-and long-term stock market prediction. in 2018 9th international conference on information and communication systems (ICICS). 2018. IEEE.
16. Ni, L., et al., Forecasting of forex time series data based on deep learning. Procedia computer science, 2019. 147: p. 647-652.
17. Wang, D., Y. Zhang, and Y. Zhao. LightGBM: an effective miRNA classification method in breast cancer patients. in Proceedings of the 2017 International Conference on Computational Biology and Bioinformatics. 2017.
18. Z. Zeng and M. Khushi, "Wavelet Denoising and Attention-based RNN-ARIMA Model to Predict Forex Price," in 2020 International Joint Conference on Neural Networks (IJCNN), 19-24 July 2020
19. Ding, Y. and I.W. Selesnick, Artifact-free wavelet denoising: Non-convex sparse regularization, convex optimization. IEEE signal processing letters, 2015. 22(9): p. 1364-1368.
20. Selesnick, I.W., et al., Simultaneous low-pass filtering and total variation denoising. IEEE Transactions on Signal Processing, 2014. 62(5): p. 1109-1124.
21. He, K., et al. Identity mappings in deep residual networks. in European conference on computer vision. 2016. Springer.
22. Zagoruyko, S. and N. Komodakis, Wide residual networks. arXiv preprint arXiv:1605.07146, 2016.
23. Akiba, T., S. Suzuki, and K. Fukuda, Extremely large minibatch sgd: Training resnet-50 on imagenet in 15 minutes. arXiv preprint arXiv:1711.04325, 2017.
24. Ju, Y., et al., A model combining convolutional neural network and LightGBM algorithm for ultra-short-term wind power forecasting. IEEE Access, 2019. 7: p. 28309-28318.
25. Min, F., et al. Fault prediction for distribution network based on CNN and LightGBM algorithm. in 2019 14th IEEE International Conference on Electronic Measurement & Instruments (ICEMI). 2019. IEEE.
26. Gong, R., et al. Acoustic scene classification by fusing LightGBM and VGG-net multichannel predictions. in Proc. IEEE AASP Challenge Detection Classification Acoust. Scenes Events. 2017.